\documentclass[aps,prd,twocolumn,showpacs,amsmath,nofootinbib]{revtex4}

\usepackage[T1]{fontenc}
\usepackage[latin1]{inputenc}
\usepackage{graphicx}
\usepackage[english]{babel}
\usepackage{graphicx}
\usepackage{bm}
\usepackage{amsmath}
\usepackage{amssymb}
\usepackage{amsfonts}
\usepackage{epsfig}
\usepackage{colordvi}
\usepackage{color}
\usepackage{epsfig}

\usepackage{subfigure,amsmath,amssymb,amsfonts, latexsym}
\usepackage[notcite,color,final]{showkeys}
\usepackage[labelsep=period]{caption}
\begin{document}

\title{ Hojman Symmetry Approach  for Scalar-Tensor Cosmology}
\author{ Mariacristina Paolella$^{1, 2,}$\footnote{E-mail: paolella@na.infn.it},
Salvatore Capozziello$^{1, 2,3}$\footnote{E-mail: capozziello@na.infn.it}}
\affiliation{$^1$Dipartimento di Fisica, Universit\`{a} di Napoli {}``Federico II'' and\\
$^2$INFN Sez. di Napoli, Compl. Univ. di Monte S. Angelo, Edificio G, Via Cinthia, I-80126, Napoli, Italy,\\
$^3$Gran Sasso Science Institute (INFN), Viale F. Crispi, 7, I-67100, L'Aquila, Italy.}

\date{\today}

\begin{abstract}
Scalar-tensor Cosmologies can be dealt under the standard of the Hojman conservation theorem that allows to fix the form of the coupling $F(\phi)$, of  the potential $V(\phi)$  and 
 to find out exact solutions for related  cosmological models. Specifically, the existence of a symmetry transformation vector for the equations of motion gives rise to a  Hojman conserved quantity on the corresponding minisuperpace    and exact solutions for the cosmic scale factor $a$ and the scalar field $\phi$ can be achieved. 
 In particular, we take advantage of the fact that minimally coupled solutions, previously obtained in the Einstein frame,  can be conformally transformed in non-minimally coupled solutions  in the Jordan frame. Some physically relevant examples are worked out. 
  \end{abstract}


\maketitle

\bibliographystyle{plain}
\maketitle

\section{Introduction}

Selecting symmetries and conserved quantities allows, in general, to reduce dynamics. As a consequence,  exact  solutions for physical systems can be  achieved. In particular, the so called \emph{Noether Symmetry Approach} is a  method to solve exactly  dynamics in many cosmological models \cite{ritis}. In other words, the existence of a Noether symmetry for the point-Lagrangians coming from a given field theory, allows  to find out first integrals for the  equations of motion. It has been adopted for several classes of  alternative theories of gravity \cite{alternative}   leading to the full solution of  dynamics \cite{noether}. 

Besides the Noether Theorem, other approaches, based on symmetries, can be adopted in order to reduce dymanics and obtain exact solutions.
For example, the Hojman conservation theorem \cite{hojman} can  provide a further method  to find exact solutions. As shown in \cite{Cap-Rosh},  the Hojman approach  does not require, a priori,  the  need for  Lagrangians and Hamiltonian functions. The symmetry vectors and the corresponding conserved quantities can be obtained by using directly the equations of motion. Specifically, unlike the Noether approach,  a point-Lagrangian from which derive  the equations of motion is not requested. The intersting fact is that  these two approaches may give rise to  different conserved quantities that,  eventually, can coincide. For example, in the case of minimally coupled scalar-tensor gravity models, the Noether symmetry exists only for exponential potential $V(\phi)$ \cite{ritis}; while  the Hojman symmetry exists for a wide range of potentials \cite{Cap-Rosh}. In general, the Hojman conserved quantities  theorem can  be  different  with respect to the Noether ones. Taking advantage of this fact and adopting  conformal transformations, it is possible to study several classes of non-minimally coupled models as we will do in this paper. 
In summary, applying the Hojman Theorem to a minimally coupled models and finding out exact solutions, by conformal transformations, we obtain exact solutions for non-minimally coupled scalar-tensor models.

The layout of the paper is the  following. In Sec. \ref{hojman ct}, we summarize the Hojman  theorem giving the details of demonstration. In Sec. \ref{scalar-tensor},    scalar-tensor theories of gravity are introduced and   the Hojman conservation theorem is applied.  First, summarizing the results achieved in \cite{Cap-Rosh}, we discuss the minimally coupled case and then, by conformal transformations, we  achieve exact solutions for general non-minimally coupled scalar tensor theories. Conclusions are drawn in Sec. \ref{concl}.

\section{The Hojman conservation theorem}\label{hojman ct}

In order to formulate  the Hojman conservation theorem, let as 
consider a system of second-order differential equations that, specifically, can be the equations of motion of a given dynamical system:
\begin{equation}
\ddot{q}^{i}=F^{i}(q^j,\dot{q}^j,t),\ \ \ \ \ \ \  i,j=1,...,n\,.
\label{neq}
\end{equation}
Dot is  the  derivative  with respect to the time. 
 A  symmetry vector $X^{i}$ for Eqs. \eqref{neq} is defined according to the
 transformation
\begin{equation}
q'^{i}=q^i+\epsilon X^i(q^j,\dot{q}^j,t),
\end{equation}
that maps solutions $q^i$ of Eqs. \eqref{neq} into solutions $q'^i$ of the same equations. Such a vector has to satisfy the equations
 \begin{equation}\label{xe}
\frac{d^2 X^i}{dt^2}-\frac{\partial F^i}{\partial q^j}X^j-\frac{\partial F^i}{\partial \dot{q}^j}\frac{dX^j}{dt}=0,
\end{equation}
where
\begin{equation}
\frac{d}{dt}=\frac{\partial}{\partial t}+\dot{q}^i\frac{\partial}{\partial q^i}+F^i\frac{\partial}{\partial \dot{q}^i}.
\end{equation}
The Hojman conservation theorem \cite{hojman} states: \\
If the  function $F$ in Eqs. \eqref{neq} satisfies the condition
\begin{equation}
\frac{\partial F^i}{\partial \dot{q}^i}=0,
\end{equation}
then
\begin{equation}\label{cq1}
Q=\frac{\partial X^i}{\partial q^i}+\frac{\partial}{\partial \dot{q}^i}\left(\frac{dX^i}{dt}\right),
\end{equation}
is a conserved quantity, that is   
\begin{equation}
\frac{dQ}{dt}=0\,.
\end{equation}
Furthermore, if $F$ satisfies
\begin{equation}\label{lambda}
\frac{\partial F^i}{\partial \dot{q}^i}=-\frac{d}{dt}\ln\gamma,
\end{equation}
where $\gamma$ is a function of $q^i$, then also
\begin{equation}\label{cq2}
Q=\frac{1}{\gamma}\frac{\partial(\gamma  X^i)}{\partial q^i}+\frac{\partial}{\partial \dot{q}^i}\left(\frac{dX^i}{dt}\right).
\end{equation}
is a conserved quantity.\\
As previously discussed in \cite{Cap-Rosh},  such a theorem can be suitably applied to dynamical equations describing cosmological models. Specifically, any cosmological model can be considered  a minisuperspace ${\mathbb{Q}}\equiv\{q_j\}$ whose dynamics is defined on the tangent space ${\mathbb{TQ}}\equiv\{q_j,\dot{q}_j\}$. If the Hojman theorem is satisfied, conserved quantities related to couplings and potentials can be find out. This feature allows the reduction of dynamics and the possibility to obtain exact solutions, as we will discuss below.


\section{ Scalar-Tensor Cosmology in the Hojman approach}\label{scalar-tensor}

In four dimensions, the action involving gravity coupled in a non-standard way with a scalar field is
\begin{equation}\label{Action_NMC}
\mathcal{A}= \int d^4x \sqrt{-g} \left[ F(\phi) R + \frac{1}{2} g^{\mu \nu} \nabla_{\mu} \phi \nabla_{\nu} \phi - V(\phi) \right],
\end{equation}
where $R$ is the Ricci scalar, $V(\phi)$ and $F(\phi)$ are generic functions describing, respectively, the potential for the field $\phi$ and
the coupling of $\phi$ with gravity.
In a flat Friedman-Robertson-Walker (FRW) space-time, the Lagrangian density for non-minimal coupled  scalar-tensor cosmology is 
\begin{equation}\label{Lagrangian_NMC}
\mathcal{L}= 6 F(\phi) a \dot{a}^2+ 6 F'(\phi) a^2 \dot{a} \dot{\phi} + \frac{1}{2} a^3 \dot{\phi}^2- a^3 V(\phi),
\end{equation}
where the prime $'$ denotes the derivative with respect to $\phi$ and the dot denotes the derivative with respect to the time t.
Eq. (\ref{Lagrangian_NMC})  is a point-like Lagrangian on the configuration space $(a,\phi)$.
The Euler-Lagrange equations relative to (\ref{Lagrangian_NMC}) are
\begin{equation}\label{EqMotion_a_NMC}
2\frac{\ddot{a}}{a}+\left(\frac{\dot{a}}{a}\right)^2+ \frac{2 F'}{F}\left(\frac{\dot{a}}{a}\right) \dot{\phi}+\frac{F'}{F}\ddot{\phi} + \left( \frac{F''}{F} -\frac{1}{4F} \right) \dot{\phi}^2- \frac{V}{2F}=0,
\end{equation}
\begin{equation}\label{EqMotion_phi_NMC}
\ddot{\phi} + 3\left(\frac{\dot{a}}{a} \right)\dot{\phi}+6 F'\left(\frac{\dot{a}}{a}\right)^2+ 6  F'\left(\frac{\ddot{a}}{a}\right)+  V'=0,
\end{equation}
which correspond respectively to the  Einstein-Friedman equation and to the Klein-Gordon equation for the FRW case.
The energy function, corresponding to the Einstein $(0,0)$ equation,  is 
\begin{equation}\begin{split}
 6 F a \dot{a}^2+  6 F' a^2 \dot{a} \dot{\phi} + \frac{1}{2} a^3 \dot{\phi}^2+ a^3 V=0. \label{energy_NMC}
\end{split}\end{equation}
It is possible to construct a conformal Lagrangian  corresponding  to a minimally coupled scalar field. This can be achieved by introducing 
the following transformations \cite{Allem_Franc}
\begin{eqnarray}\label{confor_transformation}
&\bar{a}=\sqrt{-2F(\phi)} a, \\ \nonumber
&\frac{d\bar{\phi}}{dt}=\sqrt{\frac{3F'(\phi)^2-  F(\phi)}{2  F(\phi)^2}} \frac{d\phi}{dt}, \\
& d\bar{t}=\sqrt{-2F (\phi)} dt .\nonumber
\end{eqnarray}
Under transformations (\ref{confor_transformation}), we obtain
\begin{equation}\begin{split} \label{MC_Lagrangian}
\frac{1}{\sqrt{-2F}} \mathcal{L}&=\frac{1}{\sqrt{-2F}} (6 F a \dot{a}^2+ 6 F' a^2 \dot{a} \dot{\phi} + \frac{a^3}{2}  \dot{\phi}^2- a^3 V(\phi))\\
&= - 3 \bar{a} \dot{\bar{a}}^2 +\frac{1}{2} \bar{a}^3 \dot{\bar{\phi}}^2- \bar{a}^3 \bar{V}(\bar{\phi})= \mathcal{L}_{MC},
\end{split}\end{equation}
where the dot over barred quantities means the derivative with respect to $\bar{t}$ and 
\begin{equation}\label{transf_Potential_VE}
\bar{V}(\bar{\phi}(\phi))= \frac{V(\phi)}{4F^2(\phi)}.
\end{equation}
Hence, under transformations (\ref{confor_transformation}),  the non-minimal coupled Lagrangian becomes a conformally related  
minimal coupled Lagrangian.
Following the standard terminology, the ''Einstein frame'' is the frame
with the minimal coupling and the ''Jordan frame'' is the frame with  non-minimally coupling.
This means that  for any non-minimally coupled scalar field, we may associate a unique minimally coupled scalar 
field in the  conformally related space by deriving the correct relation between the coupling and the potential as \eqref{transf_Potential_VE}.
Thus, in principle, in order to find out solutions  for the non-minimal coupled theory, we can consider solutions in the Einstein frame and transform these back
into the Jordan frame using the conformal transformations \eqref{confor_transformation}.
Such a property can be used as a solution generator in the sense that by achieving solutions in the Einstein frame through the Hojman Theorem, as in Ref.\cite{Cap-Rosh}, it is possible to derive solutions in the Jordan frame and vice-versa. In this sense, the Hojman approach is a general criterion to find out exact solutions. 


\subsection{The minimally coupled case}
Starting from the point-like Lagrangian \eqref{MC_Lagrangian} and discarding the bar over the physical quantities,  the equations of motion are given by
\begin{equation}\label{quin1}
\frac{\dot{a}^2}{a^2}=\frac{1}{3}\left[\frac{\dot{\phi}^2}{2}+V(\phi)\right],
\end{equation}
\begin{equation}\label{quin2}
\ddot{a}=\frac{a}{3}\left[V(\phi)-\dot{\phi}^2\right].
\end{equation}
The Klein-Gordon equation is 
\begin{equation}\label{quin3}
\ddot{\phi}+3\left(\frac{\dot{a}}{a}\right)\dot{\phi}+V'(\phi)=0\,.
\end{equation}
By introducing the variable $x=\ln a$, and by combining equations \eqref{quin1} and \eqref{quin2}, we find the following equation of motion \cite{Cap-Rosh}
\begin{equation}\label{qem}
\ddot{x}=-f(x) \dot{x}^2,
\end{equation}
where
\begin{equation}\label{f}
f(x)=\frac{1}{2} \phi'(x)^2.
\end{equation}
Assuming that  $a(t)$ and $\phi(t)$ are invertible functions of $t$, dynamics  can be reduced to a one dimensional motion.  Eq. \eqref{quin3} can be considered as a constraint. From Eq.\eqref{qem}, it is clear that $F(x,\dot{x})=-f(x)\dot{x}^2$, thus
\begin{equation}\label{quinl0}
\gamma(x)=\gamma_0 e^{\int f(x) dx},
\end{equation}
where $\gamma_0$ is an integration constant. Another relation is achieved   by dividing Eqs. \eqref{quin3} and \eqref{quin1}. We have
\begin{equation}\label{pot1}
\frac{V'(\phi)}{V(\phi)}=\frac{f(x)\phi'(x)-\phi''(x)-3\phi'(x)}{3-\frac{1}{2}\phi'(x)^2}.
\end{equation}
 Eq. \eqref{xe} becomes
\begin{equation}\label{quin11}
\begin{split}
&\left(f(x)\frac{\partial X}{\partial x}+f'(x)X+\frac{\partial^2 X}{\partial x^2}\right)+\dot{x}^2 f(x)^2 \frac{\partial^2 X}{\partial \dot{x}^2}-\\&\dot{x}\left(2f(x)\frac{\partial^2 X}{\partial x\partial \dot{x}}+f'(x)\frac{\partial X}{\partial \dot{x}}\right)=0\,,
\end{split}
\end{equation}
where we assumed that $X$ does not depend explicitly  on time.
If $X=X(\dot{x})$, the only solution for \eqref{quin11} is \cite{Cap-Rosh}
\begin{equation}\label{x1}
X(\dot{x})=A_0 \dot{x}^n+A_1 \dot{x},
\end{equation}
and then
\begin{equation}
f(x)=-\left(\frac{1}{nx+f_0}\right).
\end{equation}
Considering the meaning of the above variables, the generic potential with respect to $\varphi=\phi-\phi_c$ is
\begin{equation}\label{pot2}
V(\varphi)=\lambda \varphi^{\frac{4}{n}}-\frac{8\lambda}{3n^2}\varphi^{\frac{4}{n}-2},
\end{equation}
where
\begin{equation}
\lambda=3V_0\left(\frac{n^2}{8}\right)^{\frac{2}{n}}.
\end{equation}
The exact solutions  $a(t)$ and $\varphi(t)$ for the potential \eqref{pot2} are
\begin{equation}\label{exact1}
\begin{split}
&a(\bar{t})=e^{-\frac{f_0}{n}}e^{-\frac{1}{n}\left[(1-\frac{1}{n})\bar{t}\right]^{\frac{n}{n-1}}},\\&
\varphi(\bar{t})=\pm \frac{\sqrt{8}}{n}\left[(1-\frac{1}{n})\bar{t}\right]^{\frac{n}{2(n-1)}},
\end{split}
\end{equation}
where the parameter $\bar{t}$ is defined as
\begin{equation}
\bar{t}=y_0-n|Q_0|^{\frac{1}{n}}t,
\end{equation}
$Q_0$ is the Hojman conserved quantity and $\bar{t}$ can be seen as a sort conformal time ruled by $Q_0$.
In the next section we will use these results in order to find out solutions for non-minimally coupled scalar-tensor cosmologies ruled by the forms of the potential 
$V(\phi)$ and of the coupling $F(\phi)$.

\subsection{The quadratic coupling case}
\begin{figure}[htbp]
       \includegraphics[scale=0.4]{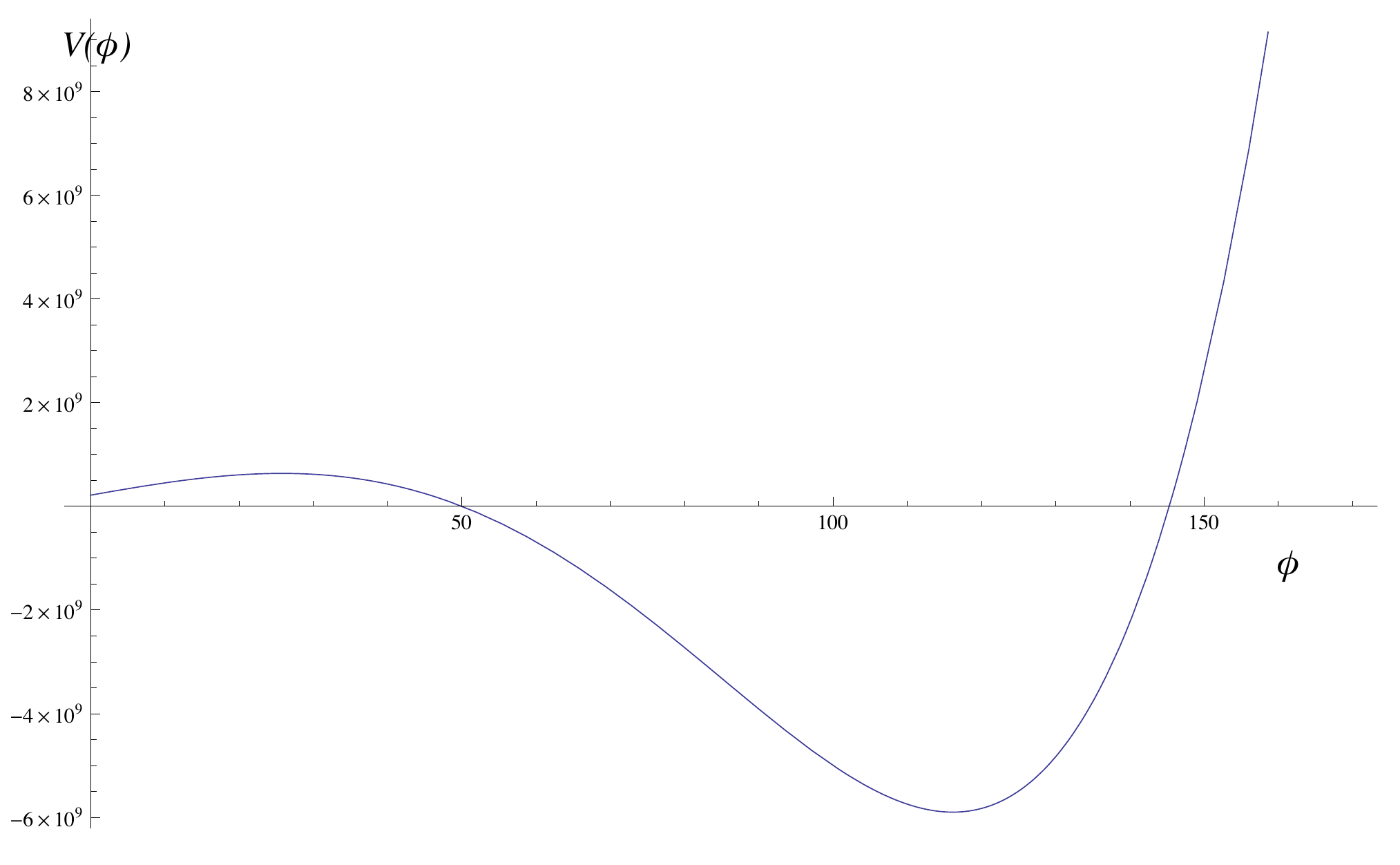}
  	   \caption{\label{Pot_caso1} Plot of the potential $V(\phi)$ for some specific values of the constants,  $k=100$, $\xi=-10$, $\phi_0=150$, $n=4/3$. }
\end{figure}

Let us start with the general case of a quadratic coupling of the form
\begin{equation}
F(\phi)= \frac{\xi}{4} (k+\phi)^2,
\end{equation}
where $k$ and $\xi$ are arbitrary constants (with $\xi<0$ in order to recover physical cases).
In this case equations \eqref{EqMotion_a_NMC} and \eqref{EqMotion_phi_NMC} in the variable $\bar{t}$ become
\begin{equation}
\frac{2 \ddot{a}}{a}+\frac{ \dot{a}^2}{a^2}+\frac{6 \dot{a}}{a} \frac{\dot{\phi}}{ (k+\phi)}+\frac{2\ddot{\phi}}{(k+\phi)}+\frac{(4\xi -1) \dot{\phi}^2}{\xi (k+\phi)^2}-\frac{4 V}{\xi^2 (k+\phi)^4}=0,
\end{equation}
\begin{equation}
\ddot{\phi} + 3\xi \frac{\ddot{a}}{a} (k+\phi) +3 (1+\xi)\frac{\dot{a} \dot{\phi}}{a}+3\xi \frac{\dot{a}^2}{a^2}(k+\phi)+\frac{\dot{\phi}^2}{k+\phi}-\frac{2 V'}{\xi (k+\phi)^2}=0.
\end{equation}
Transformations \eqref{confor_transformation} gives 
\begin{eqnarray*}
\bar{a}&=&\sqrt{-\frac{\xi}{2}} \; (k+\phi) a, \\ \nonumber
\bar{\phi}&=& \sqrt{\frac{6\xi-2}{\xi}} \;\ln[k+\phi]+ c_1, \\
d\bar{t} &=& \sqrt{-\frac{\xi}{2}} \; (k+\phi) dt. \nonumber
\end{eqnarray*}
Solutions \eqref{exact1} become
\begin{eqnarray}\label{Solution_NMC1}
  \phi(\bar{t}) &=& -k+ \phi_0  e^{\left( \frac{\sqrt{8\xi}}{n \sqrt{6\xi-2}}\left[(1-\frac{1}{n})\bar{t}\right]^{\frac{n}{2(n-1)}} \right)},\\
a(\bar{t}) &=&\sqrt{-\frac{2}{\xi}} \frac{e^{-\frac{f_0}{n}}}{\phi_0} e^{\left(-\frac{1}{n}\left[(1-\frac{1}{n})\bar{t}\right]^{\frac{n}{n-1}}-\frac{\sqrt{8\xi}}{n \sqrt{6\xi-2}}\left[(1-\frac{1}{n})\bar{t}\right]^{\frac{n}{2(n-1)}} \right)} \label{Solution_NMC2}
\end{eqnarray}
and the  potential is 
\begin{eqnarray}\label{Pot_inflation_1}
V(\phi)&=&\left(\frac{n^2 (3\xi-1)}{4 \xi}\right)^{\frac{2}{n}} \xi^2 \left( k+\phi\right)^4\left\lbrace \frac{3}{4n^2}\left[\ln\left(\frac{k+\phi}{\phi_0}\right)\right]^{\frac{4}{n}} \right.\nonumber\\
&&-\left. \frac{\xi}{n^4 (3\xi-1)} \left[\ln\left(\frac{k+\phi}{\phi_0}\right)\right]^{\frac{4}{n}-2} \right\rbrace .
\end{eqnarray}
To have an idea of what is going on, let us plot the potential  \eqref{Pot_inflation_1} in  Fig. \ref{Pot_caso1}.

\begin{figure}[!ht]
        \centering
       \includegraphics[scale=0.8]{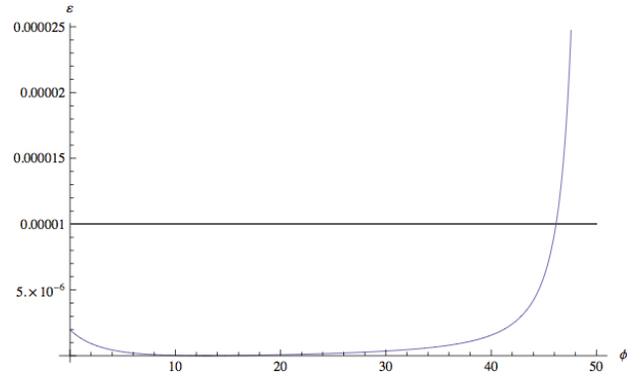}
  	   \caption{\label{epsilon_Pot_caso1} Plot of $\epsilon_{\bar{\phi}}$ with $k=100$, $\xi=-10$, $\phi_0=150$, $n=4/3$. The straight line   corresponds to $\epsilon=10^{-5}$.} 
\end{figure}

\begin{figure}[!ht]
        \centering
       \includegraphics[scale=0.8]{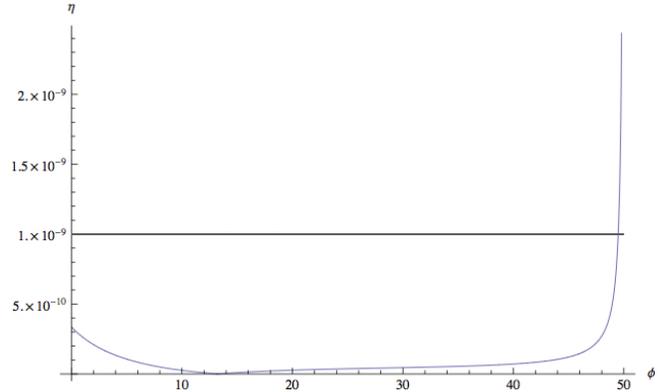}
  	   \caption{\label{eta_Pot_caso1} Plot of $\epsilon(\phi)$ with $k=100$, $\xi=-10$, $\phi_0=150$, $n=4/3$. The straight  line corresponds to 
  	   $\eta=10^{-9}$.} 
\end{figure}

The cosmology of this model  is most easily studied in the Einstein frame where the gravity sector is standard  taking advantage from the potential transformation  \eqref{transf_Potential_VE}.  
For example, the inflationary dynamics  is determined by the shape of the potential $\bar{V}(\bar{\phi})$.
It is worth noticing that  $\bar{\phi}$ (and not $\phi$)  has a canonical kinetic term. Therefore the slow-roll parameters, which 
control the first and second derivatives of the potential respectively, are
\begin{equation}
\epsilon_{\bar{\phi}}= \frac{1}{2} \left( \frac{\bar{V}_{\bar{\phi}}}{\bar{V}} \right)^2, \hspace{1cm} \eta_{\bar{\phi}} = \frac{\bar{V}_{\bar{\phi}\bar{\phi}}}{\bar{V}},
\end{equation}
where the subscript $\bar{\phi}$ means $d/d\bar{\phi}$.
In the usual way, we can formally define the first and second slow-roll parameters for the field $\phi$ that are related to the  slow roll parameters $\epsilon_{\bar{\phi}}$ and $\eta_{\bar{\phi}}$ via \cite{Topics_infl}
\begin{eqnarray}\label{slow-roll-par} 
&&\epsilon_{\bar{\phi}}= \left( \frac{d \phi}{d \bar{\phi}}\right)^{2} \epsilon_{\phi}, \\
&&\eta_{\bar{\phi}}=  \left( \frac{d \phi}{d \bar{\phi}}\right)^{2} \eta_{\phi} - \left(\frac{d^2 \phi }{d \bar{\phi}^2}\right) \sqrt{\frac{\epsilon_{\phi}}{2}}.\label{slow-roll-par1}
\end{eqnarray}
The slow-roll approximation requires that the constraints
\begin{eqnarray}
&\epsilon_{\bar{\phi}}& \ll 1 ,\\
&\left|\eta_{\bar{\phi}}\right| &\ll 1,
\end{eqnarray}
be satisfied. 
A plot of $\epsilon_{\bar{\phi}}$  is reported in Fig.\ref{epsilon_Pot_caso1} and a plot of  $\left|\eta_{\bar{\phi}} \right|$ in Fig.\ref{eta_Pot_caso1}.
The conditions $\epsilon_{\bar{\phi}} \ll 1$ and $\left|\eta_{\bar{\phi}}\right| \ll 1$ are satisfied for $0<\phi< 50$ that is the range of values of the slow-roll phase as it can be  expected from
Fig.\ref{Pot_caso1}
Thus for potential \eqref{Pot_inflation_1}, with $k=100$, $\xi=-10$, $\phi_0=150$, $n=4/3$, the conditions $\epsilon_{\bar{\phi}} \ll 1$ and 
$\left|\eta_{\bar{\phi}}\right| \ll 1$ are satisfied for $0<\phi< 50$ that is the range of values expected for the slow-roll phase as it can be seen from Fig.\ref{Pot_caso1}.

\subsection{The conformally coupled case }
\begin{figure}[!h]
        \centering
       \includegraphics[scale=0.4]{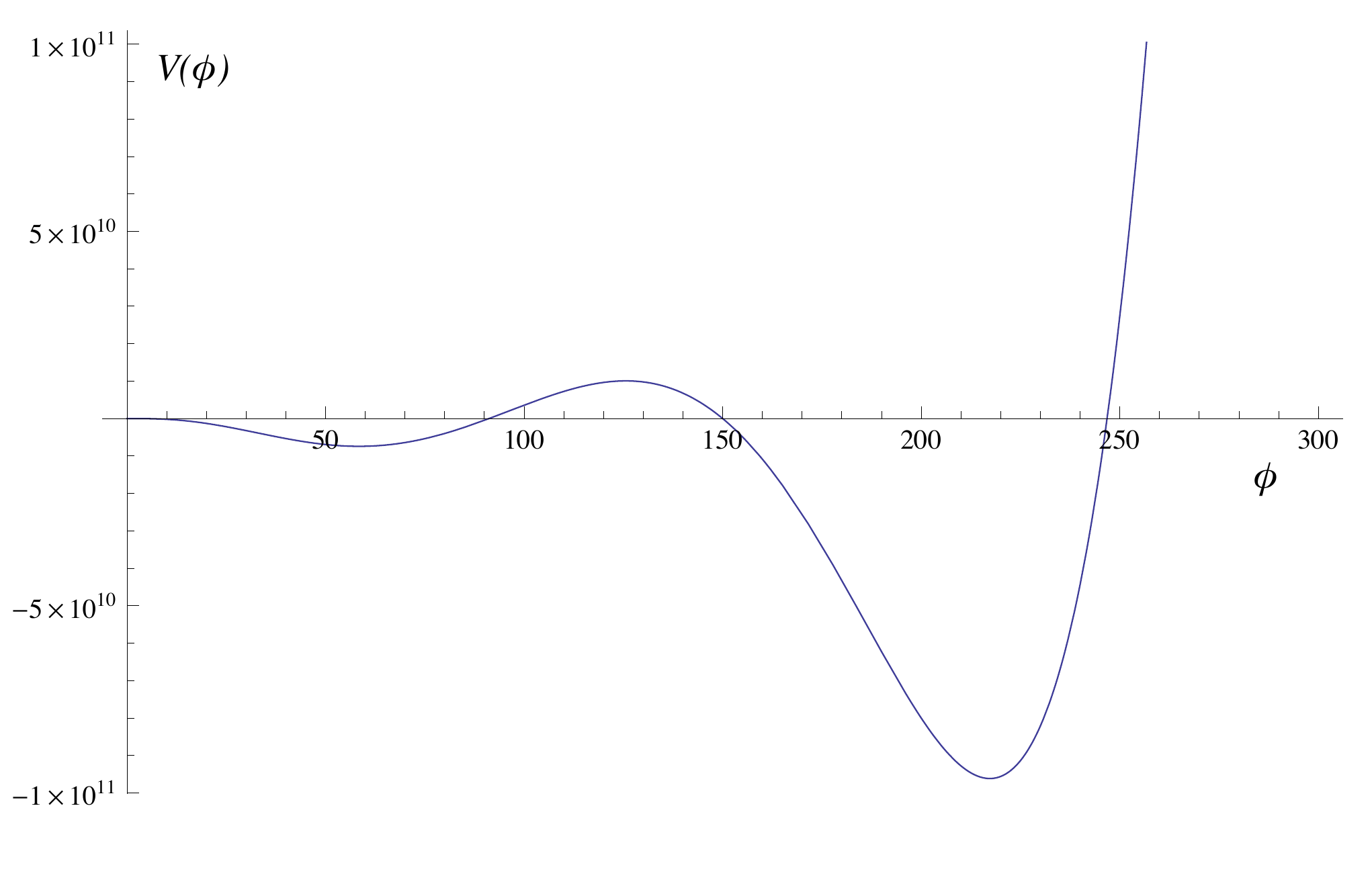}
  	   \caption{\label{plot_Pot_caso2} The potential $V(\phi)$ with $k=0$, $\xi=-10$, $\phi_0=150$, $n=4/3$ }
\end{figure}

We can  consider also  the case of conformal coupling where
\begin{equation}\label{F}
F(\phi)=\xi \phi^2, \hspace{1cm} k=0\, ,
\end{equation}
hence,
\begin{equation}
\dot{\bar{\phi}}=\sqrt{\frac{3F'^2-F}{2F^2}} \dot{\phi} = \sqrt{\frac{12 \xi - 1}{2 \xi}} \;\left( \frac{ \dot{\phi}}{\phi}\right).
\end{equation}
Integrating this relation we have
\begin{equation}
\bar{\phi}= \sqrt{\frac{12 \xi - 1}{2 \xi}} \;  \ln(\phi) + C_1 = \sqrt{c} \; \ln(\phi) + C_1.
\end{equation}
Moreover we have
\begin{equation}
\bar{a}=\sqrt{-2F (\phi)} a = \sqrt{-2 \xi} \; \phi \; a,
\end{equation}
and
\begin{equation}
d\bar{t}=\sqrt{-2F (\phi)} dt = \sqrt{-2 \xi} \; \phi \; dt.
\end{equation}
Now using solutions \eqref{exact1} for the minimally coupled case, we obtain the following solutions for the non-minimal coupling case
\begin{eqnarray}\label{Solution_NMC1}
&&\phi(\bar{t}) = \phi_0  e^{\left( \frac{\sqrt{8}}{n \sqrt{c}}\left[(1-\frac{1}{n})\bar{t}\right]^{\frac{n}{2(n-1)}} \right)},\\
&&a(\bar{t})=\frac{e^{-\frac{f_0}{n}}}{\sqrt{-2 \xi }\phi_0} e^{\left(-\frac{1}{n}\left[(1-\frac{1}{n})\bar{t}\right]^{\frac{n}{n-1}}-\frac{\sqrt{8}}{n \sqrt{c}}\left[(1-\frac{1}{n})\bar{t}\right]^{\frac{n}{2(n-1)}} \right)} . \label{Solution_NMC2}
\end{eqnarray}
The potential takes the form
\begin{eqnarray}\label{Pot_caso2}
V(\phi)&=& \xi^2 \phi^4 \left( \frac{n^2(12 \xi -1)}{16 \xi}\right)^{\frac{2}{n}} \left[  \frac{12}{n^2}  \left[\ln \left(\frac{\phi}{\phi_0}\right)\right]^{\frac{4}{n}} \right. 
\nonumber\\
&&\left. - \frac{64 \xi}{n^4(12 \xi-1)} \left[\ln \left(\frac{\phi}{\phi_0}\right) \right]^{\frac{4}{n}-2}  \right].
\end{eqnarray}
In Fig. \ref{plot_Pot_caso2}, we represent the potential for some specific values of the integration constants. Clearly the above inflationary analysis works also in this case.

\subsection{The string-like case}

Finally, let us take into account the  low energy limit of string Lagrangian, that is  
\begin{equation}\label{strin_lagr}
\mathcal{L}= \sqrt{-g} e^{-2\psi} \left[-R + 4 g^{\mu \nu} \nabla_{\mu} \psi \nabla_{\nu} \psi - W(\psi)\right],
\end{equation}
where $\psi$ is the dilaton and $W(\psi)$ is the potential which leads dynamics. Note that, due to the coupling
$e^{-2\psi}$,   modes associated with the dilaton and with  the graviton are non-minimally coupled.
Lagrangian \eqref{strin_lagr} can be immediately rewritten as a non-minimal coupled Lagrangian \eqref{Action_NMC}  if we assume the transformation \cite{Cim_Noether}
\begin{equation}\label{cond_string}
\phi(\psi)= e^{-\psi}, \hspace{0.13cm} F(\phi)=-\frac{1}{8}e^{-2\psi}, \hspace{0.13cm} V(\phi)=e^{-2\psi}W(\psi).
\end{equation}
This can be written as
\begin{equation}
\phi= - \ln \psi, \hspace{0.25cm} F(\phi)=-\frac{1}{8} \phi^2, \hspace{0.25cm} V(\phi)=\phi^2 W(\psi(\phi)).
\end{equation}
We are again in the above case with $\xi= -\frac{1}{8}$ and $\phi_0=1$.
Starting from the exact solutions \eqref{Pot_caso2},  we have that the class of potentials $W(\psi)$ which satisfy conditions \eqref{cond_string} are
\begin{equation}
W(\psi)=e^{-2 \psi} \left(\frac{5}{4}n^2\right)^{\frac{2}{n}} \left(\frac{3}{16 n^2} \psi^{\frac{4}{n}} -\frac{1}{20 n^4} \psi^{\frac{4}{n}-2} \right).
\end{equation}
and the exact  solutions are
\begin{equation}
\phi(\bar{t})= e^{\left( \frac{2}{n \sqrt{5}}\left[(1-\frac{1}{n})\bar{t}\right]^{\frac{n}{2(n-1)}} \right)},
\end{equation}
\begin{equation}
a(\bar{t})=2 e^{-\frac{f_0}{n}} e^{-\left(\frac{1}{n}\left[(1-\frac{1}{n})\bar{t}\right]^{\frac{n}{n-1}}+\frac{2}{n \sqrt{5}}\left[(1-\frac{1}{n})\bar{t}\right]^{\frac{n}{2(n-1)}} \right)}.
\end{equation}
It is straightforward to see that also in this case, the above inflationary analysis easily applies.

\section{Conclusion}\label{concl}
The Hojman Symmetry Approach can be a useful tool to find out exact solutions in dynamical systems as soon as a suitable Hojman vector is identified. Considering also conformal transformations, the method works well in both Einstein and Jordan frame with the only shrewdness that conformal transformations have to be non-singular. 
We have shown that physical solutions achieved in the Einstein frame by Hojman symmetry can be easily transformed into the Jordan frame once a relation between the potential and the coupling is found out. As examples,  we derived potentials with a clear inflationary meaning whose parameters can, in principle, be confronted with cosmological observations. In forthcoming papers, we will generalize the approaches  to other alternative theories of gravity.

\acknowledgments
MP is supported by INFN {\it iniziativa specifica} QNP. SC is supported by INFN ({\it iniziative
specifiche} TEONGRAV and QGSKY).


\end{document}